\documentclass{article}
\usepackage[cp1251]{inputenc}
\usepackage[english]{babel}
\usepackage{graphicx}
\usepackage{amsmath}
\usepackage{amssymb}
\usepackage{latexsym}
\usepackage{natbib}
\begin{document}
\begin{flushright}
Journal-Ref: Astronomy Letters, 2014, Vol. 40, No. 6, pp. 334-342
\end{flushright}

\begin{center}
\Large {\bf On the nature of the azimuthal asymmetry of protoplanetary disks observed pole-on. The case LkH$\alpha$ 101}\\

\vspace{1cm}
\large {\bf T.V.\,Demidova$^{1}$, V.P.\,Grinin$^{1,2}$}
\end{center}

\normalsize

1 - Pulkovo Astronomial Observatory of the Russian Academy of Sciences

2 - S. Petersburg State University, V.V.\,Sobolev Astronomical Institute\\

\begin{center}
e-mail:proxima1@list.ru
\end{center}
\normalsize
\begin{abstract}
The model of a protoplanetary disk of a star with a low-mass companion ($M_2:M_1 \le 0.1$), moving on a circular orbit, which slightly inclined to the disk plane ($\le 10^\circ$), is considered. The SPH-method is used to
calculate gas-dynamic flows. The motion of the
companion on the orbit leads to the inhomogeneous
distribution of the matter in the disk: there are a
cleared gap, density waves and streams of matter.
Because of the disturbances the inner part of the disk is tilted to its periphery and also do not coincides
with the orbit plane of the companion. It leads to anisotropic
illumination of the disk by the star and, as a consequence, to the appearance of large-scale inhomogeneity on the disk image: it has a bright area in the form of a ``horseshoe'' and a small shadow zone located asymmetric with respect to the line of nodes. The asymmetry of the disk image is clearly visible even if it is observed pole-on. The motion of the companion on
the orbit does not lead to the synchronous movement of the shadow and bright areas: they only make small oscillations relative to the some direction. By using the proposed model we
an fairly accurately reproduce the asymmetric image of the disk of star LkH$\alpha$ 101, observed almost pole-on. The study of such asymmetric disks opens up new opportunity for searching massive bodies around young stars.

\end{abstract}

\clearpage

\section{Introduction}
In the overwhelming majority of cases the circumstellar
disks around young stars are not resolved
through a telescope and we know about their existence
from circumstantial evidence (intrinsic linear
polarization of stars, infrared (IR) excesses, emission
line profiles, jets). Optical and IR images of
the disks were obtained for dozens of young objects
with the Hubble Space Telescope and large ground based
telescopes \citep[see, e.g.,][]{1999AJ....117.1490P, 2002ApJ...577..826T, 2005ApJ...630..958G}\footnote{The number and quality of protoplanetary disk images will increase rapidly in the coming years owing to the started operation of the ALMA interferometer.}. In many cases, the quality of these images is already high enough for them to be used to study large-scale features on the disks. Pole-on disks are of particular interest, because for such an orientation, first, the image details are seen better \citep[see, e.g.,][]{2011ApJ...729L..17H, 2012ApJ...748L..22M}. Second, and this is especially important,
the images of such disks are affected neither by the anisotropy of light scattering by dust grains nor by the self-shielding of inner disk regions near the dust sublimation zone, which give rise to an asymmetry in the images of the disks if they are inclined relative to the plane of the sky.

The existing methods of observations do not yet
allow any planets at their formation stages to be seen
in protoplanetary disks. However, their existence can
be judged from the distortions in the disk images
that they cause during their motion. Hydrodynamic
simulations performed by the SPH method \citep{1996ApJ...467L..77A, 1997MNRAS.285...33B, 2007AstL...33..594S} and by applying finite-difference methods \citep{2002A&A...387..550G, 2010ARep...54.1078K, 2010ApJ...708..485H} show that the
motion of a secondary component in a circumstellar
disk gives rise to a gap at the disk center, density
waves, gas flows, and shock waves. In the models
with a low-mass companion ($q < 0.01$), a matter-free
ring is formed near the orbital radius of the companion
instead of the gap \citep{2007A&A...471.1043D}. If the
companion's orbit is inclined relative to the plane of
the circumbinary disk, than the matter captured by
the companion from the circumbinary disk forms a disk around the binary’s primary component whose plane is inclined both relative to the plane of the outer disk and relative to the orbital plane of the companion \citep{1996MNRAS.282..597L, 1995MNRAS.274..987P, 2010AstL...36..808G,2013AstL...39...26D, 2013MNRAS.431.1320X}.
It can be seen from Fig. 1b that a global asymmetry in
the distribution of matter whose scales are considerably
larger than other hydrodynamic inhomogeneities
caused by the companion’s motion (the companion’s
accretion disk, the shock waves arising during the
collision of gas flows, etc.., which are calculated insufficiently accurately in our SPH models) arises in
the circumbinary disk.

The circumstellar disk of $\beta$ Pic, whose inner
region is inclined by several degrees relative to the
disk periphery, is an example of such a system. A
similar picture is also observed for the Herbig Ae
star CQ Tau: the inner part of its circumstellar
disk is inclined approximately by $15-20^\circ$ relative
to the disk periphery \citep{2004ApJ...613.1049E, 2006A&A...460..117D, 2008A&A...488..565C}. Observations of the Rossiter–McLaughlin effect during the transits of
planets across the disks of their host stars show that
the orbital plane does not coincide with the equatorial
plane of the central star for 40\% of these objects \citep[see,
e.g.,][]{2010A&A...524A..25T, 2012ApJ...757...18A}. Possible
causes of the appearance of planets and substellar
companions in orbits that are noncoplanar with the
circumstellar disk have been discussed in recent years
in many papers \citep{1995MNRAS.274..987P, 2011MNRAS.412.2790L, 2007ApJ...669.1298F, 2008ApJ...678..498N, 2003ApJ...597..566T, 2010MNRAS.401.1505B}, but the question about the evolution of
such binaries still remains unclear \citep{2010A&A...511A..77F, 2011A&A...530A..41B, 2013MNRAS.431.1320X}.

The perturbations in the disk caused by the orbital
motion of the companion can manifest themselves
as periodic circumstellar extinction variations when
such binaries are viewed edge-on or at a small angle
to the disk plane. The models of such binaries \citep{2010AstL...36..422D,2010AstL...36..808G} well explain the
main properties of the cyclic activity of UX Ori stars.
Perturbations in the inner regions of the disks also
affect the illumination of their peripheral regions. The
images of such disks have recently been computed
by \citet{2013AstL...39...26D} and \citet{2015A&A...579A.110R} for
the optical and submillimeter spectral ranges, respectively.
In this paper, we consider the possibility of applying
our models to explain the strong asymmetry in
the image of the circumstellar disk around the Herbig
Be star LkH$\alpha$ 101 obtained by \citet{2002ApJ...577..826T}.

\section{The method of calculation}
\subsection{Protoplanetary disk model}
\begin{figure}[!h]\begin{center}
 \makebox[0.6\textwidth]{\includegraphics[scale=1]{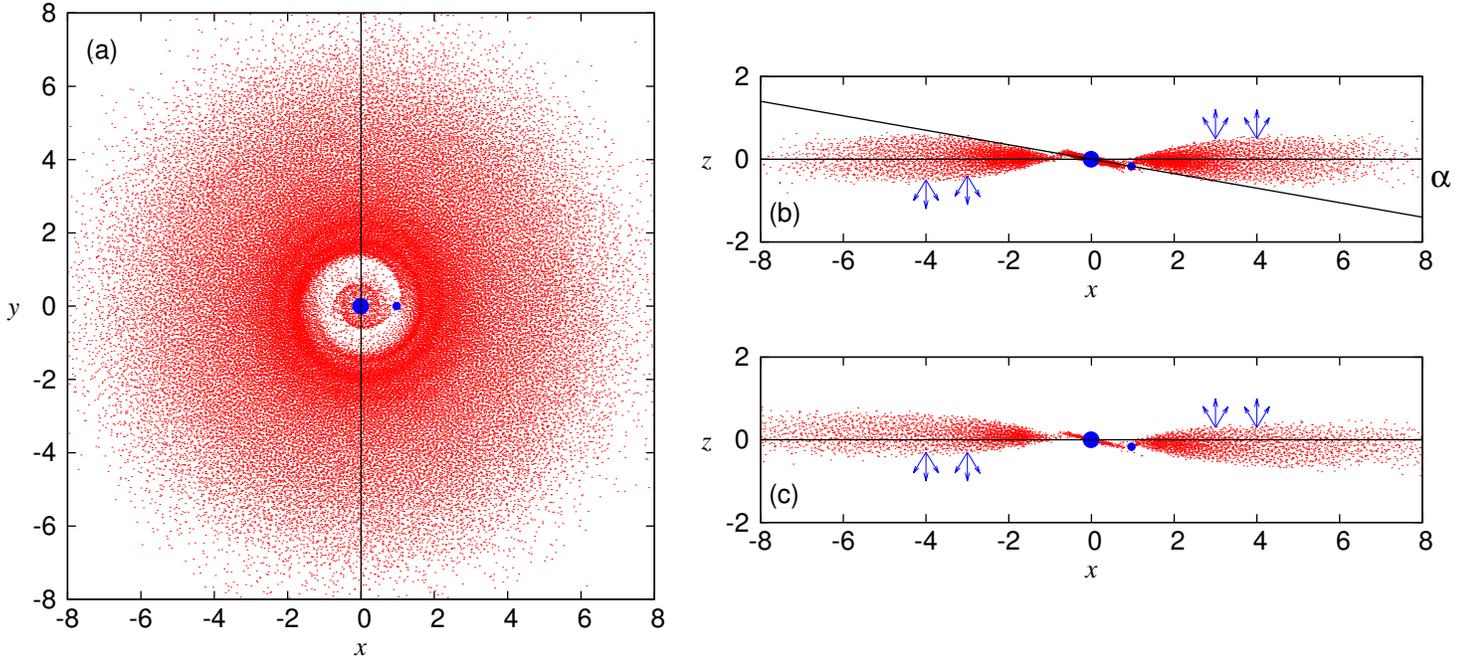}
  }
 \caption{ View of the disk after 30 binary revolutions: (a) pole-on view, (b) sections along the y axis, and (c) the same as (b) after 200 binary revolutions. The model: $q = 0.01$ and $\alpha=10^\circ$ The arrows indicate the disk regions heated by direct stellar radiation, which gives rise to a bright region on the disk. A shadow zone is formed on the opposite side.}
 \label{disk}
\end{center}
\end{figure}

We consider a star with mass $M_1$ surrounded by a
gas–dust disk in which a low-mass companion with
mass $M_2$ moves. We assume that the orbit of the
companion is circular and that its plane is inclined
to the disk midplane at a small angle $\alpha$ (Fig.~\ref{disk}b). The main parameters of the problem are the star-to-companion mass ratio $q = M_2:M_1\le 0.1$, the
orbital inclination to the disk plane $\alpha$, the effective
viscosity of the disk that was defined via the dimensionless
speed of sound $c$ expressed in fractions of
the Keplerian speed in the companion’s orbit (here,
we consider the models at $c = 0.05$), and the semi-major axis of the orbit of the companion $a$.

The calculations of the hydrodynamic flows in the
circumstellar disk caused by the companion’s motion
presented below were performed by the SPH
(Smoothed Particle Hydrodynamics) method. The
disk mass was assumed to be small compared to the
mass of the central star. This allowed the disk selfgravity
to be neglected. By analogy with the work by
\citep{1996ApJ...467L..77A}, we chose the initial
surface density distribution $\Sigma \approx r^{-1}$, the Shakura--Sunyaev viscosity parameter $\alpha_{ss}=0.03$, and the relative disk half-thickness $\delta=z/r=0.1$. In this case, the diffusion time of surface density evolution is $t_{\nu}\approx 4\cdot 10^3$ yrs. The implementation of this method is described in detail in \citealt{2007AstL...33..594S}.

At the initial time, the companion’s orbit is inclined
to the disk plane at angle $\alpha$. The time of a ``distortion'' of the inner part of the circumbinary disk due
to the companion–disk interaction can be estimated
\citep{1999MNRAS.307...79I}. At the above parameters, it
corresponds to about 10 binary revolutions, but the
``distortion'' radius can be estimated only numerically.
As our calculations show, for the given time it corresponds
to $r \approx a$.

Since the large-scale inhomogeneity arises in a
comparatively narrow disk region near the companion’s
orbit, the results of our calculations should not
depend strongly on the surface density distribution
law. This circumstance also justifies using the
isothermal approximation in hydrodynamic calculations.
In our simulations, we used $10^5$ test particles
in a region of radius $6a$. Our simulations showed that
the disk is azimuthally homogeneous at a distance
of approximately $5a$ and can be smoothly extended.
Since the sizes of protoplanetary disks can reach
several hundred AU, in our calculations we assume
the presence of an outer disk at $r > 6a$.

Our simulations showed that the binary finishes
the relaxation stage with the formation of a central
gap after 30 revolutions (see Fig.~\ref{disk}a). By this time, it loses about $5\%$ of the particles through their accretion onto the star and the companion. We assume that
the distribution of SPH particles at the given time
reflects best the distribution of matter in the simulated
protoplanetary disk region.

The model calculated in this way was smoothed
over the cells of a 3D mesh (with a step of $0.1a$). This
procedure allows the influence of random fluctuations
in the distribution of SPH particles to be reduced
while retaining all details of the flow structure.

In this paper, we do not investigate the evolution
of the inclination of the disk or orbit, because the
time of our simulations is limited ($t \le 200P$). It
is well known from the studies of similar binaries
that the circumbinary disk in binaries with a massive
companion is rapidly aligned with the binary orbit
under the action of periodic perturbations \citep[see, e.g.,][]{1999MNRAS.307...79I, 2012MNRAS.423.2597N}. In contrast, in
binaries with a low-mass companion (a planet), the
companion’s orbit is aligned with the circumbinary
disk plane \citep{2013MNRAS.431.1320X}.
The fact that the inner regions of the circumstellar
disks around $\beta$ Pic and CQ Tau are inclined relative
to the periphery suggests that such binaries can exist
for a fairly long time.

\subsection{Disk illumination by direct stellar radiation}
When calculating the surface brightness of the
protoplanetary disk, we assumed the dust to be well mixed with the gas. As in previous papers, the opacity
 $\kappa$ per gram of matter was taken to be $250$ cm$^2$ g$^{-1}$ (corresponding to the optical properties of circumstellar
dust at a wavelength of about $0.2 \mu$m; \citealt{1996A&A...311..291H}). The dust-to-gas ratio was
taken to be the same as, on average, in the interstellar
medium, $1:100$. The particle mass used in the SPH
models can be determined by two methods: first, if
the accretion rate onto the binary components $\dot M_a$ is
specified as a parameter of the problem and is compared
with the accretion rate of SPH particles, which
is calculated during our simulations. Second, the
mass of the simulated part of the disk can be specified
as a parameter of the problem. In our calculations, the
particle mass $m_d$ was $(3-6)\cdot
10^{22}$ g, corresponding
to an accretion rate $\dot M_a \approx
10^{-9} M\odot \cdot yr^{-1}$ and a mass of the simulated disk region $M_d \approx
3\cdot 10^{-6} M_\odot$. 

To calculate the illumination of the disk by direct
stellar radiation, its surface should be described
mathematically. For this purpose, the disk plane was
divided into cells in radius $r$ and azimuth $\phi$ (the
azimuthal angle is measured from the y axis that
coincides with the line of nodes in the direction of
the companion’s motion, i.e., clockwise). Having
calculated the distribution of SPH particles in the
disk for each orbital phase and given the particle
mass, we find the surface density $\Sigma(r,\phi)$ in each cell.
Hence, by assuming the matter in the disk to be
distributed according to the barometric law, we obtain the distribution of matter along $z$ at each point $r,\phi$:

\begin{equation}
\rho(r,\phi,z)= \rho_0(r,\phi)\cdot e^{-(z/h_0)^2},
\end{equation}

Here, $h_0$ is the density scale height in the disk at
point $r$: $h_0(r)=\sqrt{2}c/\Omega$, where $c$ is the speed of sound
in the disk and $\Omega$ is the Keplerian angular velocity at
this point; the density $\rho_0$ at $z=0$ is determined from
the relation

\begin{equation}
\Sigma(r,\phi) = \int_{-\infty}^{\infty}\rho(r,\phi,z)dz =
\sqrt{\pi}\,h_0(r)\,\rho_0(r,\phi),
\end{equation}

The disk surface at point $r,\phi$ was defined in such a
way that the optical depth of the disk from the top to
point $z_1$ was equal to unity ($\tau_{z_1}= \kappa\cdot\Sigma_{z_1} = 1$, where
$\Sigma_{z_1}$ is the surface density to level $z_1$) (Fig. 2). The
illumination of each cell on the disk surface by direct
stellar radiation was determined from the formula \citep{2006A&A...448..633T}.

\begin{equation}
 F_s(r,\phi,z_1) = \frac{L_{\ast}e^{-\tau(r,\phi,z_1)}}{4\pi(r^2+{z_1}^2)}\,sin(\gamma), \label{F}
\end{equation}
where $L_{\ast}$ is the star’s luminosity; $\tau(r,\phi,z_1)$ is the
optical depth of the dust layer between the star and
an arbitrary point on the disk at distance $r$ from the
symmetry axis, azimuth $\phi$, and height $z_1$, $\gamma$ is the
angle between the vector connecting the star and
the point on the disk surface and the tangent to the
disk surface at this point \citep[for details, see][]{2013AstL...39...26D}. The particles in the inner disk region, at
$r \le r_s$, where $r_s$ is the radius of the dust sublimation
zone, were not involved in our calculations of the
optical depth. We calculated $r_s$ based on Eq.~\ref{F} and
the Stefan–Boltzmann law $F =\sigma T^4$ at a silicate dust
sublimation temperature of $1500$ K.

In this paper, we consider a bright Herbig Be star
with a mass of $\sim 10M_{\odot}$ and a luminosity $L_1\approx10^3L_{\odot}$. At the mass ratio $q=0.01$, the companion’s
mass is $0.1M_{\odot}$ and it is an M-type star with a
luminosity $L_2\approx10^{-3}L_{\odot}$. We neglected the disk
illumination by the companion, because it is negligible
compared to the illumination by the central
star. However, the companion’s luminosity at the
highest point of the orbit (above the disk), at the
time when it passes near the shadow zone, can
exceed the illumination of the shadow zone. Since,
however, the companion is near the inner boundary of
the shadow zone at this time, this effect can hardly
be noticed during observations. The disk heating
through viscous dissipation (at the accretion rate
$10^{-9}M_{\odot}$ under consideration) is negligible even
near the star and was disregarded in our calculations.

\subsection{Allowance for the scattered radiation}
\begin{figure}[!h]\begin{center}
 \makebox[0.6\textwidth]{\includegraphics[scale=0.8]{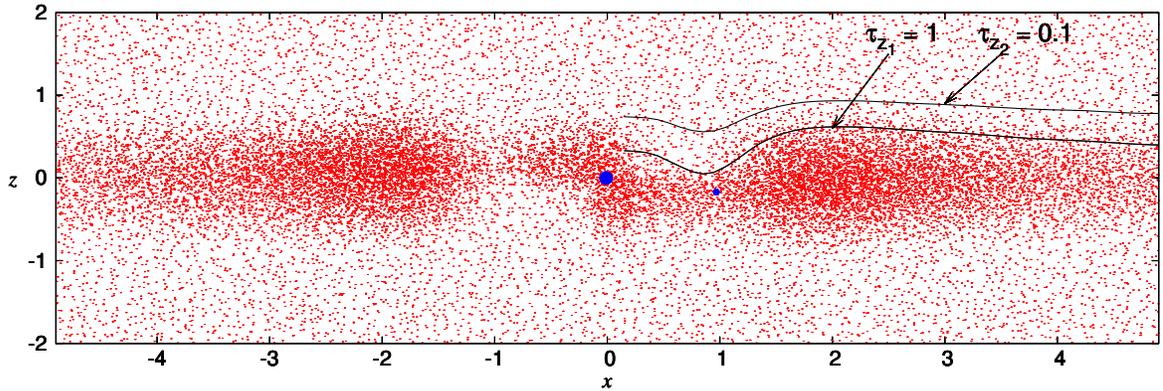}}
 \caption{ Disk section in the $z,x$ plane in model 1. The companion’s orbital radius is taken as unity. We consider the level where the vertical optical depth $\tau_{z_1} = 1$ to be the disk surface and the region between $\tau_{z_1} = 1$ and $\tau_{z_2}=0.1$ to be the scattering
layer. }
 \label{tau}
\end{center}
\end{figure}

The contribution of scattered light to the illumination
of protoplanetary disks around young stars typically
accounts for a few percent of the stellar radiation.
However, scattered light plays a significant role in the
illumination of the disk regions into which no direct
stellar radiation penetrates. In our calculations, we
assumed the scattering layer to be above the disk
surface (defined above) at such a distance $z$ from its
midplane that $0.1
\le\tau_{z} \le1$ (Fig.~\ref{tau}). The scattering
layer was divided into cells in $r$, $\phi$, and $z$. The
illumination of each volume element with coordinates
$r',\phi',z'$ in this layer was calculated from a formula
similar to (\ref{F}):

\begin{equation}
 F_s(r',\phi',z') = \frac{L_{\ast}e^{-\tau(r',\phi',z')}}{4\pi(r'^2+{z'}^2)}. \label{Fz}
\end{equation}
The disk illumination was then calculated at each
point of the surface in the single-scattering approximation
by integration over the entire volume of the scattering layer:
\begin{equation}
 F_{sc}(r,\phi,z_1) = k_{sc}\int_0^{2\pi}d\phi'\int_{r_s}^{5a}r'dr'\int_{z_1}^{z_2}
 \frac{F_s(r',\phi',z')}{4\pi|\overrightarrow{\bf{d}}|^2}
 \rho(r',\phi',z')\,f(\theta)\,e^{-\tau(r',\phi',z',r,\phi,z_1)}\,cos\,\xi\,dz'.
\label{sca}
\end{equation}

Here, $\overrightarrow{\bf{d}}$ is the vector between the center of the scattering
cell and the center of the area on the disk surface
with coordinates $r,\phi,z_1$ on which scattered light
is incident, $\xi$ is the angle between the vector $\overrightarrow{\bf{d}}$ and the
normal to the disk surface at the area center; $k_{sc}$ is the
scattering coefficient, $\tau(r',\phi',z',r,\phi,z_1)$ is the dust
optical depth on the ray running away from the scattering
cell to the area on the disk surface. We used
the Henyey–Greenstein phase function: $f(\theta) =
(1-g^2)/(1+g^2-2g\,cos\theta)^{3/2}$, where $g$ is the asymmetry
factor and $\theta$ is the scattering angle. The asymmetry
factor $g$ and the scattering albedo were taken to be $0.5$
(Mathis et al. 1977). Hence, the scattering coefficient
is $k_{sc} = 125$ cm$^2$ g$^{-1}$.

\subsection{Calculation of the disk image}
Keeping in mind the application of the results of
our calculations to the circumstellar disk of LkH$\alpha$ 101
observed in the $K$ band, we calculated the model images
for this photometric band. The total illumination
of each cell on the disk surface is $F = F_s +
F_{sc}$. By
assuming each cell on the disk surface to radiate as
a blackbody, we determined the disk temperature as
a function of  $r$ and $\phi$ from the Stefan–Boltzmann
law: $T =
(F/\sigma)^{1/4}$. The luminosity for $1$ cm$^2$ of the
disk surface was determined by taking into account
the wavelength dependence of the transmission coefficients
$f_K(\lambda)$ for the $K$ band of the \citealt{1965ApJ...141..923J} standard photometric system:

\begin{equation}
L_K(r,\phi) = \int_{\lambda} B_{\lambda}(T(r,\phi))\, f_K(\lambda)
d\lambda. \label{L}
\end{equation}
\section{Results}
\subsection{Model of the Protoplanetary Disk around LkH$\alpha$ 101}
LkH$\alpha$ 101 is a young star surrounded by a massive
circumstellar disk. Despite the large number of works
devoted to this star, its basic parameters are known
very inaccurately. For example, there is a spread from
160 \citep{1998AJ....116..890S} to 800 pc \citep{1971ApJ...169..537H}
in the estimated distance $D$ to the object. \citet{2002ApJ...577..826T} surmised (by comparing the radio and
photometric observations) that LkH$\alpha$ 101 is an early-B star with a mass $M_\ast \simeq 10-20M_\odot$ and a luminosity
$L\simeq (10-25)\cdot10^3 L_\odot$. According to their estimates,
the most probable distance to the object is $\sim 340$ pc.
On the other hand, according to \citet{2004AJ....128.1233H},
the star has a strong reddening ($A_v \simeq 10$) and a luminosity
($\ge 8\cdot10^3L_{\odot}$). They estimated its distance
to be $\sim 700$ pc.

\begin{figure}[!h]\begin{center}
 \makebox[0.6\textwidth]{\includegraphics[scale=0.3]{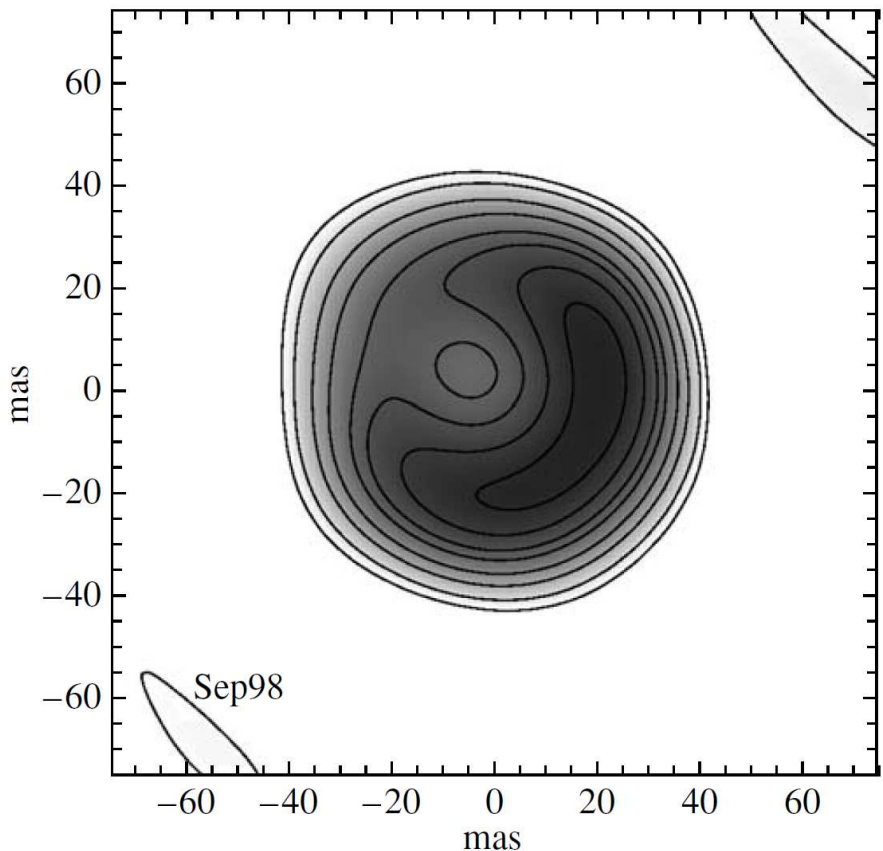}
  }
 \caption{Image of the disk around LkH$\alpha$ 101 in the CH4 filter (2.27 $\mu$m) obtained by Tuthill et al. (2002) in September
1998.}
 \label{tuthil}
\end{center}
\end{figure}

\citet{2002ApJ...577..826T} obtained a near-IR image of
the disk around Lk H$\alpha$ 101 (Fig.~\ref{tuthil}). In its central
part, the disk is asymmetric and has the shape
of a ``horseshoe''. \citet{2002ApJ...577..826T} explain this
asymmetry by assuming the disk to be inclined at a
small angle to the plane of the sky. In this case, an
observer can see part of the puffed-up inner rim in the
sublimation zone of the dust disk located behind the
star, while the frontal part of the rim will be shielded
by the disk itself. This model has two weak points:
first, the disk must be geometrically very thick and,
second, the image of a disk inclined to the plane of the
sky must have an elliptical shape. However, as can
be seen from Fig.~\ref{tuthil}, the outer boundary of the image
is described with a good accuracy by a circumference
but not by an ellipse. This suggests that the disk is
viewed nearly pole-on.

We calculated the image of the circumstellar disk
around LkH$\alpha$ 101 by assuming the star to have a low-mass
companion moving in a noncoplanar (relative to
the disk) orbit. Since the distance $D$ to the object
is known inaccurately and since there is a spread
from 160 to 800 pc, we chose two cases: 160 and 340 pc. In
the first case, (model 1, $D$ = 160 pc), the luminosity
of the central star is $L_{\ast} = 1300 L_{\odot}$, the radius of
the disk image is $R_d = 7.1$ AU ($44$ mas), the orbital
radius of the companion is $a = 4$ AU, and the radius
of the dust sublimation zone is $r_s = 2.4$ AU; in the
second case (model 2, $D$ = 340 pc), $L_{\ast} = 5900
L_{\odot}$, $R_d = 14.9$ AU, $a = 8.8$ AU, and $r_s = 5.3$ AU. In both
cases, the companion-to-star mass ratio is $q = 0.01$
and the orbital inclination is $\alpha = 10^{\circ}$.

\begin{figure}[!h]\begin{center}
 \makebox[0.6\textwidth]{\includegraphics[scale=0.4]{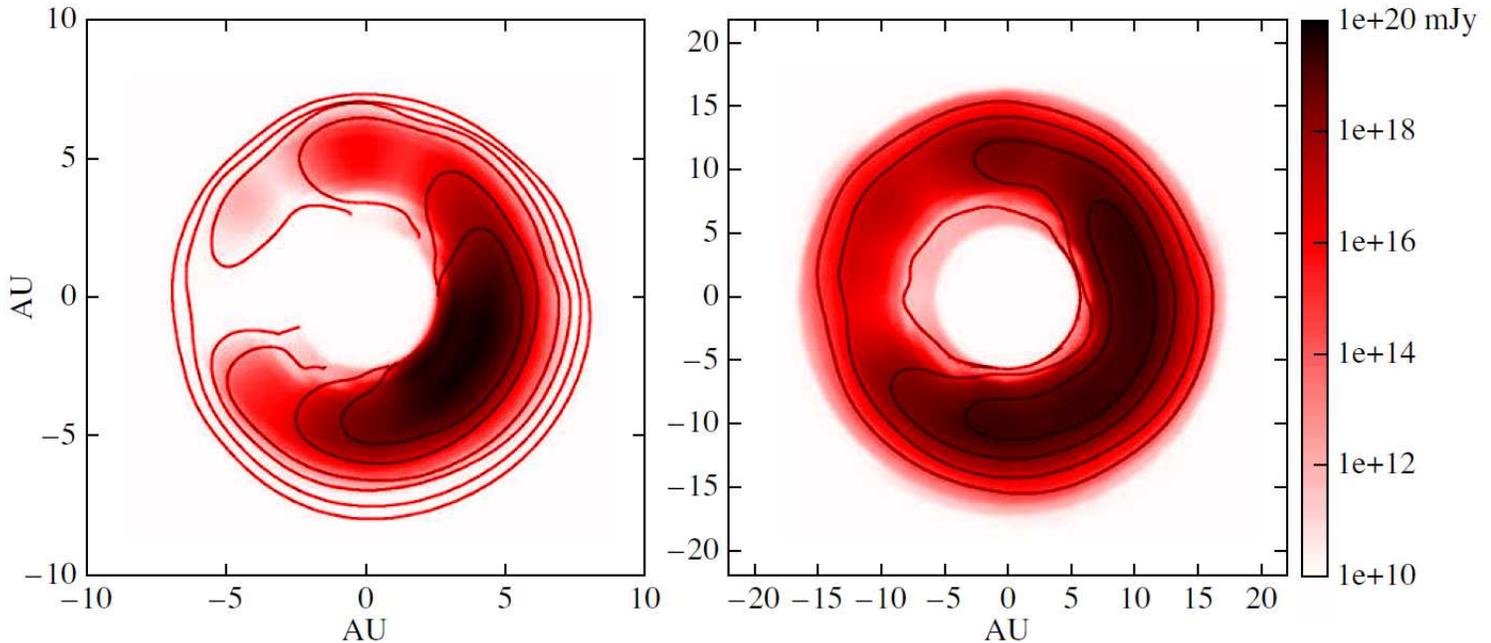}
  }
 \caption{Theoretical image of the protoplanetary disk in the K band (here and below, the color scale corresponds to the disk
surface luminosity calculated from Eq.~\ref{L}): model 1 (left panel) and model 2 (right panel). The radius of the gap in the central
part of the image is equal to the radius of the sublimation zone in the model under consideration.}
 \label{mods}
\end{center}
\end{figure}

Figure~\ref{mods} shows K-band images of the circumstellar
disk for the two models described above. In each
case, the disk is viewed pole-on, but the presence
of an inhomogeneity in the distribution of matter in
the central part of the disk leads to an asymmetry
in the disk illumination by the star. As a result, the
disk images resemble a ``horseshoe''. The presence of
a weakly illuminated region on the disk stems from
the fact that the part of the disk located within the
companion’s orbit is inclined relative to the midplane
of the remaining disk and casts a shadow.
																
\subsection{Influence of the companion’s orbital motion}

\begin{figure}[!h]\begin{center}
 \makebox[0.6\textwidth]{\includegraphics[scale=0.35]{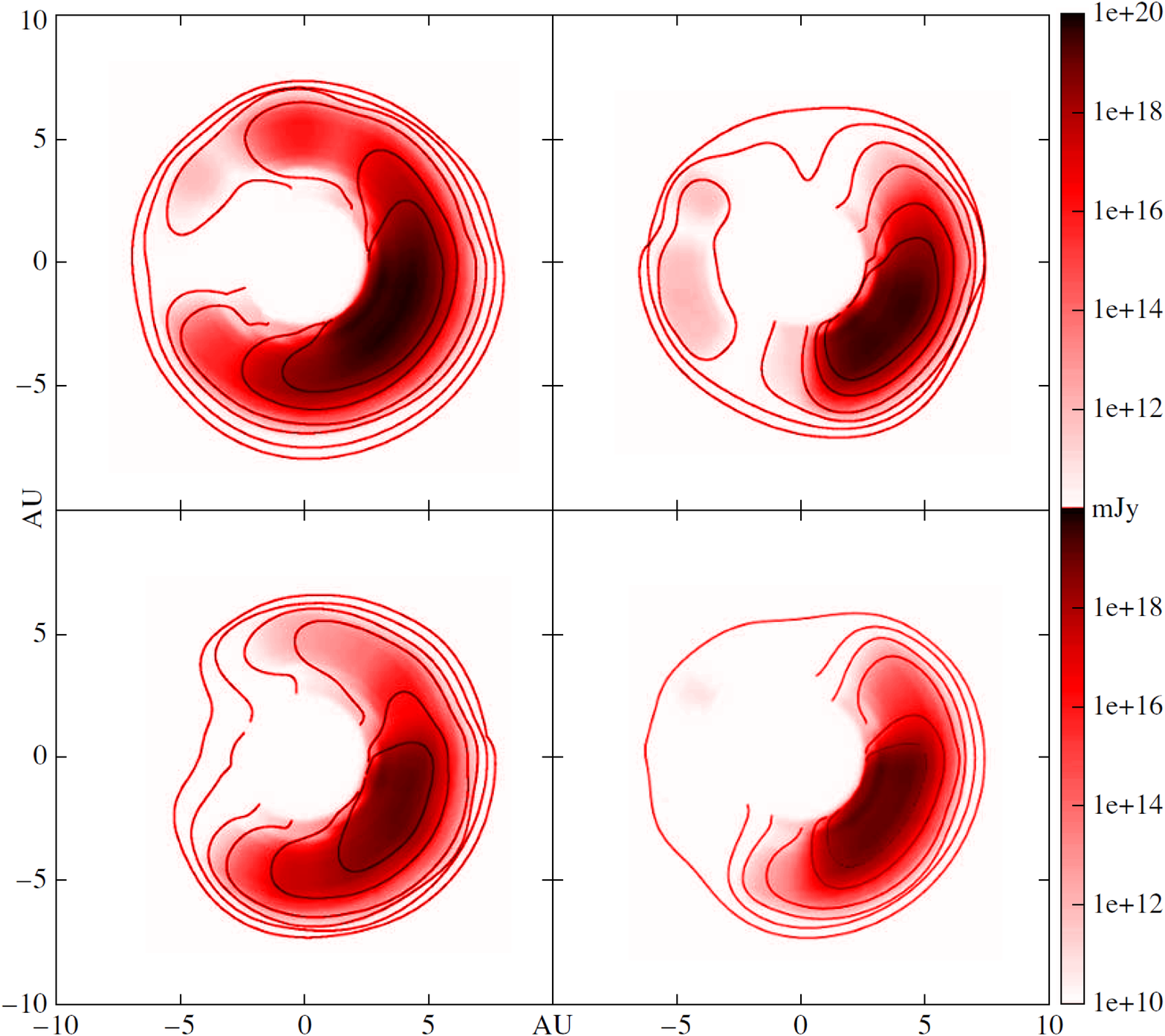}
  }
 \caption{Theoretical images of the protoplanetary disk for model 1 in the K band during one period: from left to right, from top to bottom.}
 \label{time}
\end{center}
\end{figure}

The orbital motion of the companion affects the
distribution of matter in the central part of the circumstellar
disk and leads to variations in the illumination
of its outer part. In the case under consideration,
as in our previous paper \citep{2013AstL...39...26D},
the illuminated region follows neither the rotation
of the companion nor the rotation of the disk itself
but only executes small oscillations relative some preferential direction. The sizes of the bright region
also slightly change with orbital phase, but the ``horseshoe''-shaped image remains approximately in
the same disk area (Fig.~\ref{time}). This peculiarity distinguishes
our model from other models in which the
inhomogeneities in the disk illumination move across
the disk either with the star’s rotation period (in the
model of a spotted star by \citealt{1998ApJ...506L..43W})
or with the companion’s orbital period \citep[in the model
by][]{2006A&A...448..633T}.

In the model with a noncoplanar orbit, the inner
disk region precesses \citep{1995MNRAS.274..987P,1997MNRAS.285..288L}. The precession period
depends on binary parameters. In our model, the
precession period is several hundred orbital periods.
Therefore, the influence of this factor on the disk illumination
asymmetry can be noticeable only on very
long time scales.

\subsection{Other objects}
LkH$\alpha$ 101 is one of the brightest objects in the
near-IR spectral range, because the disk is observed
around a very hot star. Similar (in shape) images of
circumstellar disks (a ``horseshoe''-shaped asymmetry
with a circular outer boundary) have recently been
observed for other stars in the mid-IR spectral range.
These include the disks around the stars HD 169142
and HD 100453 whose images at 12 $\mu$m were obtained
by \citet{2011ApJ...737...57M}. There is evidence
that HD 100453 has a companion \citep{2006A&A...445..331C}.
The Herbig Ae star HD 142527 is surrounded by a
massive gas–dust disk viewed nearly pole-on. Near-
IR \citep{2006ApJ...636L.153F} and mid-IR \citep{2006ApJ...644L.133F} observations showed that the image of the
disk around this young star has a strong asymmetry
similar to that observed for LkH$\alpha$ 101. In addition,
there is a matter-free gap in the disk between
30 and 130 AU \citep{2011A&A...528A..91V}. \citet{2006ApJ...636L.153F} hypothesize that the gap is formed by
a companion moving in an eccentric orbit
\citep[see also][]{2012ApJ...753L..38B}.

\section{Conclusions}
The calculations presented here show that the
asymmetry in the images of protoplanetary disks can
be due to an anisotropic illumination of the disk by the
central star. A nonuniform distribution of matter in
the central disk region, which affects (via extinction)
the propagation of stellar radiation, is responsible for
the azimuthally nonuniform illumination. The large-scale
inhomogeneity is caused by the perturbations
in the disk arising when the companion moves in an
orbit that is noncoplanar with the disk. As our calculations
showed, an appreciable asymmetry can be
caused by a companion whose mass is lower than that
of the central star by a factor of 100. The theoretical
disk images have a bright ``horseshoe''-shaped region
with a circular outer boundary (when viewed pole-on).
Such disk images agree well with the observational
data. The model we considered imposes no
constraints on the disk thickness and requires no disk
inclination to the plane of the sky.

\citet{2002ApJ...577..826T} presented the images of the
disk around LkH$\alpha$ 101 obtained at different dates
from December 1997 to January 2001. They slightly
change with time, but the bright ``horseshoe''-shaped
region whose position does not change is seen at all
dates of observations. In our models, the disk image
behaves similarly: the orbital motion of the companion
leads only to small oscillations in the positions
of the bright region with its overall shape retained.
This behaviour of the disk image differs fundamentally
for the models of other authors in which the bright
and dark regions on the disk follow the rotation of
a star with an inhomogeneous surface \citep{1998ApJ...506L..43W}, or the orbital motion of the companion
\citep{2006A&A...448..633T}, or the motion together
with the disk of a giant cyclone \citep{2013A&A...550L...8B}.

Apart from the global asymmetry caused by the
orbital inclination of the companion to the disk, the
companion’s motion gives rise to gas flows and shock
waves in the binary’s inner regions as well as an
accretion disk around the companion and a dusty disk
wind. Such structures can give rise to small shadow
zones on the disk that will move with the companion’s
rotation period against the background of the global
asymmetry caused by the orbital inclination to the
disk.

The question about the evolution of such binaries
remained outside the scope of this study, but, as
has been pointed out above, the existence of edge-on
circumstellar disks in which the inner region is
inclined relative to the periphery ($\beta$ Pic, CQ Tau)
suggests that the ``twisted'' disk can exist for a fairly
long time. Our calculations showed that the part of
the disk located within the companion’s orbit slowly
precesses. However, the precession rate is too low for
this effect to be detectable during observations.

It should be emphasized that the model we considered
is applicable only to optically thick disks (i.e.,
in the optical and near-IR ranges). As the radiation
wavelength increases, the contrast between the dark
(shadow) and bright regions of the disk decreases,
but the shape of the asymmetry remains as before.
In the submillimeter range, the disk is transparent
to radiation and a bright region arises at the location
of the shadow zone (the part of the disk heated
on the opposite side is transparent, see Fig.~\ref{disk}b).
The differences between the disk images in different
spectral ranges described above are of interest from
the viewpoint of observationally testing the proposed
model and deserve a further study.

{\bf Acknowledgments.} We are grateful to N.Ya. Sotnikova, who provided
the software package for hydrodynamic model calculations.
This work was supported by a grant from the
St. Petersburg Committee for Science and Higher
School, Program P21 of the Presidium of the Russian
Academy of Sciences, and grant NSh-1625.2012.2.
\\

\bibliography{biblio}
\end{document}